\documentclass[12pt]{article}
\usepackage{rotating}
\usepackage{epsfig,amssymb}
\input{epsf}

\def\laq{\raise 0.4 ex \hbox{$<$}\kern -0.8 em\lower 0.62 ex\hbox{$\sim$}}
\def\gaq{\raise 0.4 ex \hbox{$>$}\kern -0.7 em\lower 0.62 ex\hbox{$\sim$}}

\def\beq{\begin{equation}}
\def\eeq{\end{equation}}
\def\beqa{\begin{eqnarray}}
\def\eeqa{\end{eqnarray}}

\def\AJ{{\it Ap. J.} }

\def\CQG{{\it Class. Quantum Gravity} }

\def\MNRAS{{\it Mon. Not. R. Ast. Soc.} }

\def\NP{{\it Nucl. Phys.} }
\def\PL{{\it Phys. Lett.} }

\def\PR{{\it Phys. Rev.} }

 \def\frac#1#2{{\textstyle{{#1}\over {#2}}}}
 \def\lsim{\mathrel{\rlap{\lower4pt\hbox{\hskip1pt$\sim$}}
    \raise1pt\hbox{$<$}}} \def\gsim{\mathrel{\rlap{\lower4pt\hbox{\hskip1pt$\sim$}}
    \raise1pt\hbox{$>$}}}
\def\sqr#1#2{{\vcenter{\vbox{\hrule height.#2pt
         \hbox{\vrule width.#2pt height#1pt \kern#1pt
         \vrule width.#2pt}
         \hrule height.#2pt}}}}

\def\gappeq{\mathrel{\rlap {\raise.5ex\hbox{$>$}} {\lower.5ex\hbox{$\sim$}}}}

\def\lappeq{\mathrel{\rlap{\raise.5ex\hbox{$<$}}
{\lower.5ex\hbox{$\sim$}}}}

\begin{document}
\pagestyle{plain}

\begin{flushright}
DF/IST--2.2007\\
1st March 2007
\end{flushright}
\vspace{15mm}

\begin{center}

{\Large\bf A Curvature Principle for the interaction between universes$^*$}

\vspace*{1.0cm}

Orfeu Bertolami$^{**}$ \\
\vspace*{0.5cm}
{Instituto Superior T\'ecnico, Departamento de F\'\i sica, \\
Av. Rovisco Pais, 1049-001 Lisboa, Portugal}\\

\vspace*{2.0cm}
\end{center}

\begin{abstract}

\noindent
We propose a Curvature Principle to describe the dynamics of interacting universes in
a multi-universe scenario and show, in the context of a simplified model, how interaction drives
the cosmological constant of one of the universes toward a vanishingly small
value. We also conjecture on how the proposed Curvature Principle suggests a solution for the entropy paradox of a
universe where the cosmological constant vanishes.

\end{abstract}

\vfill
\noindent\underline{\hskip 140pt}\\[4pt]
{$^*$ Essay selected for an honorable mention by the Gravity Research Foundation, 2007} \\
\noindent
{$^{**}$ Also at Centro de F\'\i sica dos Plasmas, Instituto Superior T\'ecnico,
Lisbon} \\
{E-mail address: orfeu@cosmos.ist.utl.pt}

\newpage
\section{Introduction}
\label{sec:intro}

The fundamental underlying principle of Einstein's theory of
general relativity is the connection between curvature and matter-energy.
This relationship, as established by Einstein's field equations, is
consistent with all experimental evidence to considerable accuracy (see e.g. \cite{Will05,BPT06} for
reviews), however, there are a number
of reasons, theoretical and experimental, to question the general theory of relativity as the
ultimate description of gravity.

From the theoretical side, difficulties arise from the strong gravitational field
regime, associated to the existence of spacetime singularities, and the
cosmological constant problem. Quantization of gravity is likely
to bring relevant insights to overcome these problems,
however, despite the success of quantum gauge field theories in describing the
electromagnetic, weak, and strong interactions, the recipes they suggest to
describe gravity at the quantum level are not sufficient to achieve a fully consistent formulation.
At a more fundamental level, one can say that, the two cornerstones of modern physics,
quantum mechanics and general relativity, are not compatible with each other.

On the experimental front, recent cosmological
observations leads one to conclude that the standard Big Bang scenario of the origin
and evolution of the universe requires the introduction of ``invisible''
fields, as most of the energy content of the Universe
seems to be composed of presently unknown components, dark matter and dark energy,
which permeate much, if not all spacetime.
Nevertheless, general relativity allows for quite detailed
predictions, for instance, of nucleosynthesis
yields and the properties of the Microwave Background Radiation, and hence one can use the theory to
establish the specific properties of the missing links. In fact, it is widely believed
that one has to admit new fundamental scalar fields to achieve a fully consistent picture of
universe's evolution. Indeed, scalar fields are required to obtain a successful period of inflation
(see e.g. \cite{Olive90} for a review),
to account for the late accelerated expansion of universe, either through, for instance, a quintessence scalar field
(see e.g. \cite{Copeland06} for a review) or via the Chaplygin gas model \cite{Bento02}, and
in the case of some candidates for dark matter, either self-interacting
\cite{Bento01} or not \cite{Nunes99}.

Furthermore, given that Einstein's theory does not provide the most
general way to establish the spacetime metric, it is natural to consider additional fields,
especially scalar fields. Of particular relevance, are the scalar-tensor
theories of gravity as they mimic a great number of
unification models. The  graviton-dilaton system in
string/M-theory can, for instance, can be seen as an specific scalar-tensor
theory of gravity. For an updated discussion of the implications for these theories
of the latest high-resolution measurements of the PPN
parameters $\beta$ and $\gamma$, see e.g. \cite{DEMG06} and references therein.

However, likewise general relativity, none of its extensions seem to warrant a fully consistent description of
our universe given the huge discrepancy between the observed value of the cosmological constant
and the one arising from the Standard Model. Many solutions have been proposed to tackle this major difficulty
(see e.g. \cite{Weinberg}) and it has been remarked that it should admit a solution along the lines of
the strong CP problem have \cite{Wilczek}, which might be implemented in the
context of a S-modular invariant $N=1$ supergravity quantum cosmological model in
a closed homogeneous and isotropic spacetime \cite{BSchiappa}.
Nevertheless, none of the mechanisms
proposed to solve the problem are quite consistent (see e.g. Refs. \cite{Carroll} for recent
reviews). Actually, even in the context of string theory, the most studied quantum gravity approach,
no satisfactory solution has ever been advanced \cite{Witten}, even though more
recently, it has been argued that a solution arises if the ``landscape'' of vacua of the theory is interpreted
as a multi-universe (see e.g. \cite{ST06} and references therein). In this approach, each vacuum configuration in
the multitude of about $10^{500}$ vacua of the theory \cite{BoussoPol} is regarded as a distinct universe,
from which follows that some criteria is required for the selection of the suitable choice for the vacuum
of our universe. Anthropic arguments \cite{Susskind} and quantum cosmological considerations \cite{HMersini} have
been suggested for this vacuum selection, and hence, as
a meta-theory of initial conditions. These proposals are a relevant contribution to a better understanding
of the problem, although may not be the last word as
it should be kept in mind that a non-perturbative formulation of string theory is largely unknown \cite{Polchinski}.

In this work we propose a new mechanism for achieving a vacuum with a vanishingly small cosmological constant. It is
based on the assumption that, likewise the dynamics of matter in the physical spacetime, vacua dynamics and
evolution should emerge from a Curvature Principle that sets the way how different components of a multi-universe
interact. Actually, the interaction between different universes has already been
suggested as a possible way to obtain a vanishing cosmological constant \cite{Linde}. In quantum
cosmology, in some attempts to solve the cosmological constant problem, a ``third quantization'' has been
suggested where universes could be created and destroyed through quantum transitions \cite{Coleman}. Our approach
follows the same
logics, but it assumes that the relevant quantities to consider are the curvature invariants of each universe
of the multi-universe network. It is suggested that these invariants evolve in a ``meta cosmological time'' scale, so to
relax the curvature of one universe and place it into another. This is the core of the proposed curvature principle.

\section{The Model} \label{sec:model}

Let us consider universes whose spaces that are globally hyperbolic and satisfy the
weak and strong energy conditions. Furthermore, we assume,
for simplicity, that the topology of the components of the
multi-universe is trivial and
that the overall geometrical characterization of each universe, labeled with the
index, i, is fully specified by the curvature invariant
$I_{i} = R_{\mu \nu \lambda \sigma}^{i}  R^{\mu \nu \lambda \sigma}_{i}$,
where $ R_{\mu \nu \lambda \sigma}^{i}$ is the Riemann tensor of each universe.
This invariant stands out in comparison to other known curvature invariants
as it is not a total derivative in $4$
dimensions, as is the case of Euler densities, and it is sensitive to the
presence of singularities and of a non-vanishing the vacuum
energy. For sure, the dynamics of each universe is described by its Einstein equation, but its vacuum also
depends on the interaction with another universes through the Curvature Principle that is suggested as follows.
Since we are concerned with the vacuum of each universe, which is supposed to be homogeneous,
isotropic and Lorentz invariant\footnote{The connection between the cosmological constant
and Lorentz invariance has been discussed in different contexts in Refs. \cite{OB97, BC06}.}, then:

\beq
R_{\mu \nu \lambda \sigma}^{i} = k_i [g_{\mu \lambda}^i g_{\nu \sigma}^i - g_{\mu \sigma}^i g_{\nu \lambda}^i]~~,
\label{eq:riemann}
\eeq
for a constant $k_i$,
which correspond to de Sitter (dS), anti-de Sitter (AdS) or Minkowski spaces
whether $k_i <0$, $k_i > 0$ or $k_i=0$. From the vacuum Einstein equation with a cosmological constant, $\Lambda_i$,
it follows that $\Lambda_i = 3 k_i (1-N/2)$, where $N$ is the number of spacetime dimensions.
Moreover, it is clear that the curvature invariant is proportional to the square of the cosmological constant.
For sure, the chosen curvature invariant is unsuitable to distinguish between AdS and dS spaces; however, this
is not of particular relevance for our discussion as we will be primarily concerned with dS spaces. Notice that
dS and AdS spaces are related by analytic continuation and that invariance under complex transformations has been
proposed a possible way to solve the cosmological constant problem \cite{Hooft06}.

Let us now propose a scheme for the evolution of the curvature invariants in a ``meta cosmic time'', $T$, a time that is
related to the dynamics of interacting universes. The relation between this time and the usual
cosmic time will be discussed in a while. Clearly, one must endow the vacuum of each universe with a dynamics.
For simplicity,
let us consider only two universes and assume that their evolution is determined by the ``Lagrangian'' function:

\beq
L = {1 \over 2} \left({dI_1 \over dT}\right)^2 + {1 \over 2} \left({dI_2 \over dT}\right)^2 - V(I_1, I_2)~~,
\label{eq:lagrange}
\eeq
where $V(I_1, I_2)$ is a ``potential'' function. Of course, the construction of the potential
function is at the very heart of the proposed mechanism. Clearly, what is needed are well defined minima for
the curvature invariants and an interaction term. A fairly generic possibility is the following:

\beq
V(I_1, I_2) = \alpha_1 I_1 + \beta_1 (I_1 - I_1^{(0)})^2 + \alpha_2 I_2 + \beta_2 (I_2 - I_2^{(0)})^2 - \gamma I_1I_2~~,
\label{eq:potential}
\eeq
where $I_1^{(0)}$ and $I_2^{(0)}$ correspond to the minimal values of the curvature invariants for
universes $1$ and $2$, respectively. All coefficients of the
potential are positive and $V(I_1, I_2) \ge 0$.
One can easily see that if $\alpha_1 =\alpha_2=0$, then $I_1^{(0)} = I_2^{(0)}= 0$. A more interesting possibility arises when,
say $\alpha_2=0$, but $\alpha_1 \not=0$ as in this case $I_1 = I_1^{(0)} = 0$, however $I_2 = I_2^{(0)} \not=0$,
that is to say that the interaction
between the two universes drives the curvature invariant of universe $1$ toward a vanishing cosmological constant,
while toward a non-vanishing value for
the universe $2$. Notice that the condition of minima requires that $4 \beta_1 \beta_2 > \gamma^2$.

It is easy to see that from the ``integral of motion''

\beq
E= H \equiv {1 \over 2} \left({dI_1 \over dT}\right)^2 + {1 \over 2} \left({dI_2 \over dT}\right)^2 + V(I_1, I_2) ~~,
\eeq
that $E=0$ and that one can obtain a suitable Lyapunov function, $Ly=-H$, from which one can show that the minimum for the case
where  $\alpha_2=0$, but $\alpha_1 \not=0$ are attractors of the autonomous dynamical system associated to the motion
of $I_1$ and $I_2$.

Of course, the solutions of the equations of motion, $I_1=I_1(T)$ and $I_2=I_2(T)$, correspond to extrema of the
``action'' that can be constructed from the ``Lagrangian'' function, Eq. (\ref{eq:potential}).
However, this is not sufficient to fix the values of the
curvature invariants. This is done thanks a suitable potential. Our choice seems plausible, but it is clearly
an {\it ad hoc} one. At the present stage of our knowledge on can only conjecture whether the suggested
Curvature Principle can be accommodated within the framework of a fundamental quantum gravity proposal.

Let us now discuss the typical time of change of the curvature invariants. Even though
the cosmological constant problem is an ubiquitous problem it arises more acutely during the cosmological
phase transitions when the relevant effective potential changes from a situation where the order parameter
vanishes to a situation where it is non-vanishing, generating in the process, a large cosmological constant. Hence, the
typical time scale of change of the curvature invariants must be of order of the characteristic time of change
of the cosmological phase transitions order parameter, that is to say that it is typically a microscopic time
interval. Furthermore, given that one aims to set the overall geometrical features of each universe via the change of the
curvature invariants, then it must not differ significantly of the Hubble characteristic time of each universe at the transition, that is:

\beq
T_i \equiv \left({1 \over I_i} {d I_i \over dT}\right)^{-1} \lappeq H_i^{-1}~~.
\label{swift}
\eeq

That is to say that while a phase transition takes place, interaction between different universes change so
to cancel the curvature invariant associated with the vacuum of one of the universes. It is conceivable that
the vanishing of the cosmological constant of a given universe after multiple phase transitions might require
considering and modeling the interaction among various ``nearby'' universes.

A general point that one can make from the proposed mechanism is that according to Eq. (\ref{swift}), it
is likely that the observed accelerated expansion of our universe is not due to some residual cosmological
constant. Even though cosmological data do not exclude this possibility, supernovae data, baryon acoustic oscillations,
microwave background radiation shift parameter and topological
considerations are consistent with alternative sources for the accelerated expansion rather than the cosmological constant \cite{Data}.

\section{Discussion and Outlook}

The cosmological constant problem challenges our knowledge about the vacuum of the theories that we regard as fundamental.
It has also been shown to resist all attempts of a solution
that rely on a single universe framework. Given, that a multi-universe complex has been
recently discussed, most particularly, in the framework of the vacua landscape of string theory, it is natural
to ask whether these universes might interact. On the other hand, it is clear that a vacua theory, i.e. the non-perturbative
formulation of the fundamental quantum gravity theory, is needed to fully understand the cosmological constant problem and,
it is then just logical that an important ingredient of this formulation involves the interaction between different
universes.

In this work we have proposed a scheme involving the interaction of different universes
through their curvature invariants. The interaction is such that at vacuum it can
drive one of the invariants to vanish. The main ingredient of the proposal is the
interaction between different universes.
This is the main difference from
other schemes that constrain curvature invariants and metric related functions.
Indeed, in the unimodular gravity proposal, for instance, the determinant of the metric is non-dynamical and the
cosmological constant is shown to be an integration constant \cite{Einstein19,UW89,OB95}; in the limiting
curvature proposal, the value of the curvature invariants are bound from above so to avoid singularities
\cite{Markov82,BMS93}. Another, curvature-type principle arises in the context of the field theory of closed strings,
where minimal area metrics are proposed to solve the problem of generating all Riemann surfaces \cite{Zwiebach92}.
We suggest that this interaction can be modeled via the
curvature invariant of each universe depicted by the square of the Riemann tensor, which is sensitive the vacuum state
and is determined in each universe by the Einstein equation. If a Curvature Principle like the one suggested here
could bring some insight on the vacua properties, it would be a transcendental vindication of Einstein's genius. For sure,
if this type of principle can arise in the the context of some fundamental quantum gravity theory, it would be an important
validation. On the other hand, it is conceivable that a theory of initial conditions and interactions
between different universes lie beyond the realm of the fundamental theory and, if so, the cosmological constant might
be the only guidance available to unravel the ultimate nature of our world.

Before drawing this work to an end, let us point out that a possible implication of the
proposed Curvature Principle concerns the entropy paradox of our universe. Indeed,
the fact that our universe seems to have emerged from a singular state suggests that its initial entropy is much larger
than the one that can be accounted at the present. Penrose had suggested that the problem could be understood through the assignment
of entropy to the gravitational field through a curvature invariant, the square of the Weyl tensor \cite{Penrose}. We propose,
instead, that one should consider the curvature invariant $I$ we have been discussing. In fact, from another universe point
of view, our universe
can be regarded as a Schwarzschild black hole with its mass all
concentrated in some point and, hence $I=48 M^2 r^{-6}$, where $r$ is the
horizon's radius and $M$ its mass. We have used units where $G=\hbar=c=1$. Therefore, if the entropy scales with the volume,
then $S \sim r^3 \sim I^{-1/2}$; if the entropy scales according to the holographic principle, suitable for AdS spaces
\cite{Fischler,Bousso}, then
$S \sim r^2 \sim I^{-1/3}$. In either case, one finds that $S \rightarrow 0$ in the early universe and, $S \rightarrow \infty$
when $\Lambda \rightarrow 0$. As discussed, the latter corresponds to the universe at late time,
which is consistent with the generalized second principle of thermodynamics for our universe.

\vspace{0.5cm}

{\bf Acknowledgments~~}

\noindent
This work is partially supported by Funda\c{c}\~ao para a Ci\^encia e a
Tecnologia (Portugal) under the project POCI/FIS/56093/2004.



\bibliographystyle{unstr}

\end{document}